\documentclass[12pt]{article}

\usepackage{scicite,bm,amsmath,graphicx}

\usepackage[pdftex]{color}

\usepackage{times}

\newenvironment{sciabstract}{%
\begin{quote} \bf}
{\end{quote}}

\renewcommand{\b}{\textbf}

\newcounter{lastnote}

\title{Braiding photonic topological zero modes}

\author
{Jiho Noh,$^{1}$ Thomas Schuster,$^{2}$ Thomas Iadecola,$^{3}$ Sheng Huang,$^{4}$\\ Mohan Wang,$^{4}$ Kevin P. Chen$^{4}$, Claudio Chamon,$^{5}$\\ Mikael C.~Rechtsman$^{1}$\\
\\
\normalsize{$^{1}$Department of Physics, The Pennsylvania State University}\\
\normalsize{University Park, Pennsylvania 16802, USA}\\
\normalsize{$^{2}$Department of Physics, University of California, Berkeley, California 94720, USA}\\
\normalsize{$^{3}$Joint Quantum Institute and Condensed Matter Theory Center,}\\
\normalsize{Department of Physics, University of Maryland, College Park, Maryland 20742, USA}\\
\normalsize{$^{4}$Department of Electrical and Computer Engineering,}\\
\normalsize{University of Pittsburgh, Pittsburgh, PA 15261, USA}\\
\normalsize{$^{5}$Physics Department, Boston University, Boston, Massachusetts 02215, USA}\\
\\
}

\date{}

\begin{document} 


\baselineskip24pt


\maketitle


\begin{sciabstract} 
A remarkable property of quantum mechanics in two-dimensional (2D) space is
its ability to support ``anyons," particles that are neither fermions nor bosons. 
Theory predicts that these exotic excitations can be realized as bound states
confined near topological defects, like Majorana zero modes trapped in vortices
in topological superconductors.  Intriguingly, in the simplest cases the nontrivial
phase that arises when such defects are ``braided" around one another is not
intrinsically quantum mechanical; rather, it can be viewed as a manifestation of
the geometric (Pancharatnam-Berry) phase in wave mechanics, enabling the
simulation of such phenomena in classical systems. Here we report the first
experimental measurement in any system, quantum or classical, of the geometric
phase due to such a braiding process. These measurements are obtained using
an interferometer constructed from highly tunable 2D arrays of photonic waveguides. 
Our results introduce photonic lattices as a versatile playground for the experimental 
study of topological defects and their braiding, complementing ongoing efforts in
solid-state systems and cold atomic gases.
\end{sciabstract}



Quantum particles were long thought to fall into one of two classes: fermions or bosons.
The two classes are distinguished by the effect on the many-particle wavefunction of
exchanging any pair of such particles: for fermions, the wavefunction acquires a minus
sign, whereas for bosons it does not.  It was later understood~\cite{LeinaasMyrheim}
that, while these are indeed the only two possibilities in three or more spatial dimensions,
the 2D case admits a remarkable generalization of the concept of bosons
and fermions known as anyons~\cite{Wilczek1,Wilczek2}.
Mathematically, it is useful to think of anyons in (2+1)-dimensional spacetime, where their exchanges are
viewed as ``braids"~\cite{Frohlich,Witten,NayakReview} in which the particles' worldlines
are wound around one another. In general, exchanging two anyons (i.e. performing a single braid)
can yield any phase between 0 (bosons) and $\pi$ (fermions)~\cite{LeinaasMyrheim}.
Even more strikingly, in certain scenarios the phase acquired after performing a series of exchanges can depend
on the order in which the exchanges occurred~\cite{Frohlich,Witten}. Aside from its fundamental scientific
importance, this ``non-Abelian" braiding of anyons has attracted substantial interest in the condensed
matter physics and quantum information communities as it can be used as a basis
for robust quantum information processing~\cite{Freedman,KitaevComputation,NayakReview}.

Anyons have been theoretically predicted to arise in a variety
of topological phases of matter, for example in fractional quantum Hall systems~\cite{Halperin,Arovas},
where they can exist as deconfined quasiparticle excitations, and in topological superconductors,
where Majorana bound states nucleate at topological defects such as domain walls and vortices.
Despite substantial effort and progress~\cite{Camino,An,Willett,Nakamura,Aasen,Karzig}, as of yet there has been no conclusive experimental evidence of anyonic braiding, Abelian or non-Abelian.

Perhaps counterintuitively, the non-Abelian phases acquired upon
braiding Majorana bound states in superconductors can be understood
from the viewpoint of noninteracting particles, wherein the
single-particle Schr\"odinger equation describes particles as
waves. In this picture, the braiding phases are geometric
phases~\cite{Pancharatnam,Berry1,Berry2} arising from the
the adiabatic variation of the phase texture of the bound-state
wavefunctions as the vortices are wound around one another~\cite{Ivanov}.
Therefore, non-Abelian braiding in its simplest
incarnation can be viewed as a universal wave phenomenon accessible
beyond electronic systems.  This implies that it can be realized
experimentally in the context of photonics, where a wide range of
topological phenomena have been predicted and observed recently
\cite{HaldaneRaghu, Soljacic2009, RechtsmanFloquet, Hafezi2013, RevModPhys.91.015006}.  In fact, non-Abelian gauge fields have recently been observed in a photonic device~\cite{yang2019synthesis}, hinting at the possibility of photonic braiding.  In general, photonic topological devices (as well as those in other bosonic systems) are expected to have entirely complementary applications to
their condensed matter analogues.


In this paper we report on the first measurement of the geometric phase arising from braiding topological defects in an array of photonic waveguides, fabricated using the femtosecond direct laser-writing technique \cite{Davis,Szameit2010}. Following the theoretical proposal of Ref.~\cite{Iadecola2016}, we realize topological defects as vortices in a vector field that encodes the displacements of each waveguide in the array.  The vortices realized in our experiment bind localized topological modes whose single-particle wavefunctions are identical to those of Majorana bound states in a 2D topological superconductor.\footnote{Note however that the model we realize is in symmetry class BDI, while the topological superconductor is in class D~\cite{Chiu}; thus the two systems have important fundamental differences.} Consequently, at the noninteracting level the effect of braiding these vortices is the same as what is expected for Majorana bound states. We experimentally realize such vortices in the waveguide array and measure the effect of braiding one such vortex with a second that resides outside the array at an effectively infinite distance from the first. In order to eliminate the effect of dynamical phases, a 180$^\circ$ braiding operation is performed in two adjacent arrays, each with a vortex at its center. If the sense of rotation of the two arrays is the same (opposite), the relative phase at the core is found to be $0$ ($\pi$).  This observation matches the theoretically predicted geometric phase, providing a clear signature of braiding.


We arrange the waveguides in a near-honeycomb lattice, with each waveguide displaced from its honeycomb position $\b{r}=(x,y)$ by an $\b{r}$- and $z$-dependent amount $\b{u}_{\b{r}}(z)$. 
The diffraction of light through this waveguide array is governed by the paraxial wave equation
\begin{equation}
i\partial_{z}\psi(\textbf{r},z)=-\frac{1}{2k_{0}}\nabla^{2}_{\textbf{r}}\psi(\textbf{r},z)-\frac{k_{0}\Delta n(\textbf{r})}{n_{0}}\psi(\textbf{r},z),
\label{eq:propagation}
\end{equation}
where $\psi(\textbf{r},z)$ is the envelope function of the electric field $\textbf{E}(\textbf{r},z)=\psi(\textbf{r},z)e^{i(k_{0}z-\omega t)}\hat{x}$, $k_{0}=2\pi n_{0}/\lambda$ is the wavenumber within the medium, $\lambda$ is the wavelength of light, $\nabla^{2}_{\textbf{r}}$ is the Laplacian in the transverse $(x,y)$ plane, and $\omega=2\pi c/\lambda$. Here $n_{0}$ is the refractive index of the
ambient glass and $\Delta n$ is the refractive index relative to $n_{0},$ which acts as the potential in Eq.~\eqref{eq:propagation}.
Since the displacements are small compared to the lattice spacing $a$ ($|\b{u}| \leq .25 a$) and vary slowly in the $z$-direction, the propagation of light through the waveguide array can also be described by a coupled-mode (i.e. tight-binding) equation:
\begin{equation}\label{eq_paraxial}
i \partial_z c_{\b{r}}(z) = \sum_{\langle \b{r}' \rangle} [ t + \delta t_{\b{r},\b{r}'}(z) ] c_{\b{r}'}(z)
\end{equation}
where $c_{\b{r}}(z)$ denotes the amplitude of light in waveguide $\b{r}$, and the $z$-dependent hopping modification $\delta t_{\b{r},\b{r}'}$ arises from the change in waveguide separation due to the displacements $\b{u}_{\b{r}}, \b{u}_{\b{r}'}$.
Identical to electrons in graphene, the two sublattices of the honeycomb lattice give rise to two energy bands in the spectrum, which touch at two gapless ``Dirac'' points at crystal momenta $\b{K}_{\pm} = (\pm 4\pi / (3\sqrt{3}a) , 0)$.

\begin{figure}
\centering
\includegraphics[width=.5\textwidth,page=1]{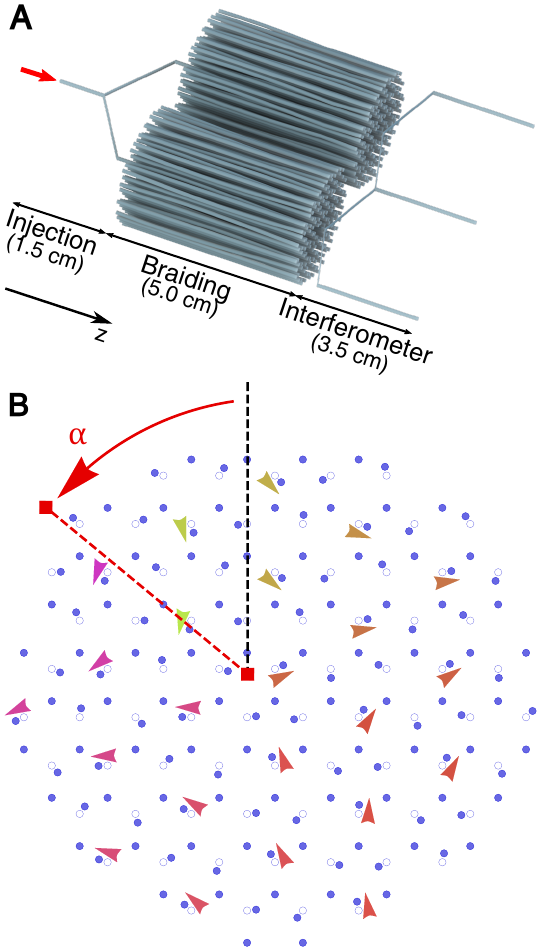}
\caption{
(A) Schematic of the waveguide array (dimensions not to scale).  The experiment consists of three stages, occurring in succession in the propagation direction, $z$.  First light is injected (indicated by the red arrow), split, and coupled into the vortex cores of the two lattices, each of which then undergoes a braiding operation.  In the interferometer stage, light is extracted from the vortex cores and combined in a single waveguide at the center.  The left and right arms are used to measure the intensity of light from each lattice after braiding.
(B) Depiction of one of the two lattices of the waveguide array at a fixed $z$-slice during the braiding stage. The waveguides (filled circles) are displaced from their honeycomb positions (empty circles) at an angle equal to the phase of the Kekul\'e order parameter $\Delta_\b{r}(z)$ (arrows, drawn parallel to the displacements and colored according to their orientation). The order parameter contains a vortex of charge $-1$ (central red square) near the lattice center. The overall offset $\alpha$ of the order parameter's phase is varied as a function of $z$ and can be interpreted as the angle between the central anti-vortex and a fictitious vortex `at infinity' (outer red square) that resides outside the waveguide array.
\label{fig1}}
\end{figure}

As shown in Fig.~\ref{fig1}, we choose the displacements $\b{u}_{\b{r}}=(u^x_{\b{r}},u^y_{\b{r}})$ corresponding to a Kekul\'e distortion~\cite{Hou,Iadecola2016} controlled by the complex order parameter $\Delta_\b{r}(z)$.
The magnitude of the order parameter determines the displacement amplitude, while its phase controls the displacement angle according to $\text{arg}(u^x_{\b{r}}+i u^y_{\b{r}}) = \b{K}_{+} \cdot \b{r} + \text{arg}(\Delta_\b{r})$.
When $\Delta_\b{r}(z)$ is constant in space, the Kekul\'e distortion couples the two Dirac cones, leading to a band gap proportional to $| \Delta |$ in the energy spectrum.

Our work concerns order parameters that vary both throughout the lattice and in the $z$-direction. 
In particular, we focus on vortex configurations of the order parameter, $\Delta_{\b{r}} = | \Delta | \exp(i [\alpha + q_{v}\,\text{arg}(\b{r} - \b{r}_v)])$, where the phase of $\Delta_{\b{r}}$ winds by $2\pi q_{v}$ about the vortex center $\b{r}_v$ for a vortex of charge $q_{v}$.
For $q_{v}=\pm 1$, such vortices are known to bind a single mid-gap photonic mode, localized near the vortex core \cite{JackiwRossi,Hou,Iadecola2016}.
The presence of this mode is protected by the nontrivial topology of the vortex configuration, as well as the time-reversal and chiral symmetries\footnote{Here, chiral symmetry arises in the effective coupled mode equation due to the bipartite coupling between the two sublattices of the honeycomb lattice, similar to graphene \cite{Semenoff}.} of the system \cite{TeoKane}.
A system with multiple vortices carries a bound mode at each vortex.
Away from the vortex centers, the order parameter varies slowly compared to the lattice scale, and the system remains gapped.
This opens up the possibility of \emph{braiding} the vortex modes as functions of $z$: if braiding is executed adiabatically in $z$, light bound in a vortex mode will not disperse into the gapped bulk.

Braiding has two effects on a vortex: moving its center $\b{r}_v$, and altering the offset $\alpha$ in the local order-parameter phase.
The latter arises due to the inherent nonlocality of vortices---as two vortices braid, each one changes its location in the space-dependent phase field of the other.
The first vortex therefore experiences an effective offset $\alpha + q_{v,2}\,\text{arg}(\b{r}_{v,1} - \b{r}_{v,2})$, which increases or decreases depending on the handedness of the braid and the charge of the second vortex.
The effect of braiding on light trapped in the vortex mode is captured by the offset-dependence of the vortex-mode wavefunction $c^v_\b{r}(\b{r}_v,\alpha)$.
Specifically, the wavefunction is \emph{double-valued} in the offset,
\begin{equation}\label{eq_doublevalue}
c^v_\b{r}(\b{r}_v,\alpha + 2\pi) = -c^v_\b{r}(\b{r}_v,\alpha),
\end{equation}
signifying that light trapped in the vortex mode gains a geometric phase of $\pi$ (i.e. a minus sign) after a full $2\pi$ braid.

The double-valued form of the wavefunction is reminiscent of Majorana wavefunctions at vortices in $p + ip$ superconductors~\cite{Ivanov, ReadGreen} or at superconductor-TI interfaces~\cite{FuKane}.
Indeed, due to this double-valuedness, the photonic zero modes in our system will gain the \emph{same} non-Abelian geometric phases as Majorana zero modes in solid state systems upon braiding~\cite{Iadecola2016}.
Crucially, this phase is gained by the \emph{bosonic} wavefunction describing photons in our system---not the Majorana wavefunction describing electrons in a superconductor---leading to a distinct, reduced class of operations realizable via braiding~\cite{Iadecola2016}.

In this work, we provide a robust verification of the geometric phase $\pi$ gained by the photonic vortex modes under a $2\pi$ rotation of the order parameter $\alpha$.
We detect this phase by performing two `on-chip' interferometry experiments, in which we interfere light from two vortex modes that have undergone different rotations of $\alpha$.
In each experiment, we fabricate a waveguide array containing two disconnected Kekul\'e-distorted honeycomb lattices, which constitute the `left' and `right' arms of an interferometer [see Fig.~\ref{fig1}(A)].
The left and right lattices are initially identical and contain a single vortex in the Kekul\'e order parameter.
As $z$ increases, the offset $\alpha$ of each lattice's order parameter is monotonically increased, or decreased, by $\pi$.
The first experiment serves as a control, with $\alpha\to\alpha+\pi$ in an identical sense of rotation in both lattices. 
In the language of braiding, this increase of $\alpha$ can be interpreted as braiding a vortex `at infinity' counterclockwise around each lattice by an angle $\pi$ [as indicated in Fig.~\ref{fig1}(B)]. 
Since the left and right lattices are identical throughout, light trapped in the left vortex mode should gain the same phase as light trapped in the right (both dynamical and geometric phases), and they should interfere constructively after braiding.

The second experiment serves to detect the geometric phase from a $2\pi$ rotation of $\alpha$.
Here the left lattice undergoes a rotation $\alpha\to\alpha+\pi$, while the right undergoes $\alpha\to\alpha-\pi$.
This is analogous to braiding a vortex at infinity $180^\circ$ counterclockwise about the left lattice, but clockwise about the right [\cite{Supplementary}, Movies S1 and S2].
Although the final left and right lattices are identical, the paths they have traversed differ by a $2\pi$ rotation of $\alpha$, and the relative phase gained by light propagating through the two vortex modes will differ according to the geometric phase of this process.
We note that chiral symmetry fixes the energy of the vortex mode to occur at precisely the middle of the band gap~\cite{JackiwRossi, Hou, Iadecola2016}, such that, despite their different $z$-evolutions, light in the left and right lattices is expected to gain the same dynamical phase during braiding.


The entire experimental waveguide array consists of three stages, occurring in succession as a function of $z$ [see Fig.~\ref{fig1}(A)].
The total length of the sample is 10 cm and the radii of the major and minor axes of the waveguides are 5.35 and 3.5 $\mu$m, respectively.
The first stage prepares the light to be coupled in-phase into the vortex modes of the left and right lattices.
This is done using an `on-chip' Y-junction beam-splitter~\cite{Izutsu}. We first precisely couple light from a tunable-wavelength laser through a lens-tipped fiber into a single waveguide at the input facet of the sample.
We then split this waveguide into two in a symmetric manner, such that both waveguides trap light of nearly equal amplitude and phase.
The two waveguides are subsequently guided to the center of the vortex cores of the left and right lattices. 

The middle stage contains the braiding operation, and begins by abruptly initializing all other waveguides in each lattice.
Light traveling in the waveguides of the first stage enters each lattice through a single waveguide near each vortex core, setting the initial condition for the braiding process. 
The portion of the input light that overlaps with the vortex mode remains localized in the mode, while the rest diffracts throughout the lattice.
After initialization, the waveguide displacements are smoothly varied in $z$ to produce the desired change in order parameter for the experiment being performed.
We emphasize that the individual waveguides move little during this process: each waveguide only undergoes rotation of its displacement angle from the undistorted honeycomb position.
%

\begin{figure}
\centering
\includegraphics[width=.725\textwidth,page=1]{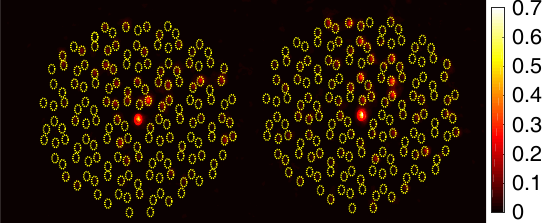}
\caption{Diffracted light measured at the output facet of a waveguide array device containing only the injection and braiding stages.  Light stays largely confined to the vortex core, while excess light that does not overlap with the vortex mode diffracts throughout the array. A color bar is presented on the right, which indicates the light intensity at the output facet on a relative scale.\label{fig2}}
\end{figure}

In Fig.~\ref{fig2}, we show the experimental output of a waveguide array containing only these first two stages. 
As expected, a high fraction of the light is localized near each vortex core, indicating that the initialization stage succeeded and the braiding process was performed sufficiently adiabatically.
We also observe a nonzero intensity in nearly all waveguides of both lattices, indicating that light not bound to the vortex modes has diffracted throughout each lattice.

To detect the phase of the vortex mode light after braiding, we add a third, interferometric stage to the waveguide array~\cite{Izutsu}, depicted in Fig.~\ref{fig1}(A).
The left and right lattices are abruptly terminated except for one waveguide each, chosen for its high overlap with the final vortex-mode wavefunction.
Light localized in the left and right vortex modes continues to propagate through the remaining two waveguides. The intensity and phase of these waveguides' light are proportional to those of the respective vortex modes after braiding.
In order to read out the waveguides' relative phase, we first split each waveguide symmetrically into two, similar to the first stage. 
The two innermost waveguides of the left-right pair are then combined into a single waveguide, while the two outermost remain separate.
The waveguide combination performs the interferometry: if the two input arms are in phase, they will excite the symmetric bound mode of the combined waveguide; if they are out of phase, they will not overlap the bound mode and thus the light will diffract away. 
The combined waveguide intensity $I_C$ thus indicates the intensity of in-phase light.
The outermost left and right waveguide intensities $I_L$ and $I_R$ measure the waveguide intensities before combination.

\begin{figure}
	\centering
	\includegraphics[width=.725\textwidth,page=1]{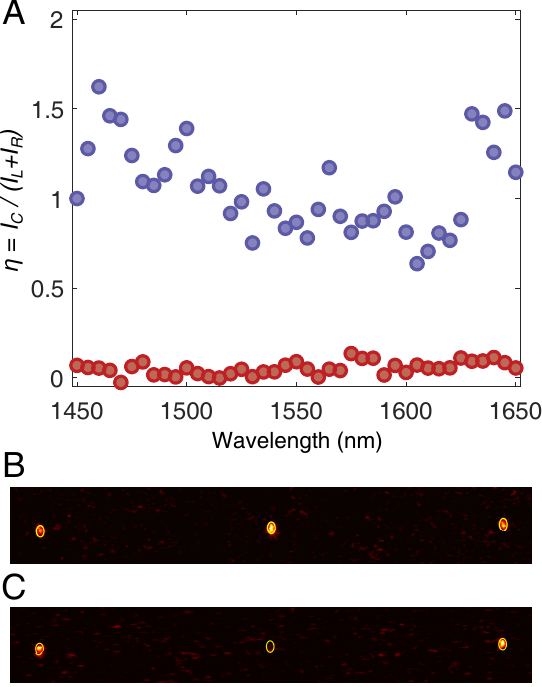}
	\caption{(A) The experimentally observed contrast when the two lattices are braided in the same direction (blue) or in opposite directions (red).  The geometric phase $\pi$ picked up by light near the vortex core in the latter case is evidenced by the near-zero contrast due to destructive interference. (B, C) Diffracted light measured at the output facet when the two lattices are braided in the same and opposite directions, respectively.   \label{fig3}}
\end{figure}

Results of both experiments including the final stage are shown in Fig.~\ref{fig3}. 
As anticipated, braiding the left and right vortices in the same direction results in a combined waveguide with higher intensity than the individual left and right waveguides, indicating constructive interference.  
On the other hand, braiding the left and right vortices in opposite directions results in nearly zero combined waveguide intensity, indicating near-complete destructive interference and therefore a relative phase of $\pi$ between the left and right vortex modes.
To quantify these results, we define the \emph{contrast} $\eta$ as the ratio between the combined waveguide intensity and the sum of the individual left and right waveguide intensities, $\eta = I_C / (I_L + I_R)$. 
In the ideal case, the contrast achieves the upper bound $1$ for perfect constructive interference and the lower bound $|I_L - I_R|/(I_L + I_R)$ for perfect destructive interference.
As plotted in Fig.~\ref{fig3}(A), the experimentally observed contrast is indeed near $1$ when braiding is performed in the same direction for both lattices,
and is close to zero (the lower bound for symmetric intensities, $I_L = I_R$) when it is performed in opposite directions.

We verify that this phase difference arises as a geometric, rather than a dynamical, phase by demonstrating that the interference is insensitive to the wavelength of light used.
The wavelength sets the `time' scale in the paraxial equation, Eq.~(\ref{eq_paraxial}), so that different-wavelength light acquires different dynamical phases during propagation~\cite{Guglielmon2018}.
If the relative phase between the left and right waveguides arose as a dynamical phase, we would expect it to oscillate as the wavelength of light was varied rather than being quantized at $\pi$. 
Instead, as shown in Fig.~\ref{fig3}, we observe quite consistent, non-oscillatory values of $\eta$ for both experiments over a large range of wavelengths, 1450--1650 nm.  The principal sources of error/noise are interference from imperfectly in-coupled light in the braiding stage as well as diffracted light in the interferometric stage.  We note that the random error is significantly lower in the sample in which we observe the $\pi$ phase associated with braiding [red points in Fig.~\ref{fig3}(A)] compared to the `control' sample in which we do not [blue points in Fig.~\ref{fig3}(A)].  
The measured $\eta$ values when the two vortices are braided in equal and opposite directions are 1.055$\,\pm\,$0.247 and 0.053$\,\pm\,$0.037, respectively.  
This indicates that the relative phase is fixed to $\pi$ and supports its identification as a geometric phase.

In conclusion, we have used a photonic lattice of evanescently-coupled waveguides to directly measure the braiding of vortices.  The $\pi$ phase that we observe here directly implies that non-Abelian braiding operations can be carried out using similar ideas in this and many other platforms that are governed by classical wave equations.  We expect this work to motivate the exploration of braiding physics in interacting bosonic systems (via optical nonlinearity or mediation by Rydberg atoms, for example), as well as the search for applications beyond robust quantum information processing.


\bibliographystyle{Science}

\section*{Acknowledgments}

M.C.R. and J.N. acknowledge the National Science Foundation under award number ECCS-1509546 and the Penn State MRSEC, Center for Nanoscale Science, under award number NSF DMR-1420620.  M.C.R. acknowledges Office of Naval Research under award number N00014-18-1-2595 and the Packard foundation under fellowship number 2017-66821.
T.S. acknowledges support from the National Science Foundation Graduate Research Fellowship Program under Grant No. DGE 1752814.
T. I. acknowledges support from the Laboratory for Physical Sciences, Microsoft, and a JQI postdoctoral fellowship.
C.C. is supported by DOE Grant No.~DE-FG02-06ER46316.


\section*{Competing interests}
The authors declare no competing interest.


\newpage
\section*{Supplementary materials}
\subsection*{Movie S1 and S2: Movies of the entire waveguide array throughout the braiding stage.}
Movies of the two waveguide lattices throughout the braiding stage, with each frame representing a constant-$z$ slice of the waveguide array. The waveguides (filled circles) are displaced from their honeycomb positions (empty circles) at an angle equal to the phase of the Kekul\'e order parameter $\Delta_\b{r}(z)$ (arrows, drawn parallel to the displacements and colored according to their orientation). The order parameter in each lattice contains a vortex of charge $-1$ (central red square) near the lattice center. The overall offset $\alpha$ of the order parameter's phase in each lattice is varied as a function of $z$ and can be interpreted as the angle between the central anti-vortex and a fictitious vortex `at infinity' (outer red square) that resides outside the waveguide array. 
\textbf{(Movie S1)} In the first experiment $\alpha\to\alpha+\pi$ in both lattices, corresponding to braiding a vortex at infinity counterclockwise about each lattice. 
\textbf{(Movie S2)} In the second experiment, $\alpha\to\alpha+\pi$ in the lattice on the left, and $\alpha\to\alpha-\pi$ in the lattice on the right, corresponding to braiding a vortex at infinity counterclockwise about the left lattice but clockwise about the right lattice. The final configurations of the two lattices differ by a phase $2\pi$ and are identical.

\clearpage

\vfill

\end{document}